# Observation of a Hidden Hole-Like Band Approaching the Fermi Level in K-Doped Iron Selenide Superconductor

Masanori Sunagawa[1*], Kensei Terashima[1,2], Takahiro Hamada[1], Hirokazu Fujiwara[1], Tetsushi Fukura[1], Aya Takeda[1], Masashi Tanaka[3], Hiroyuki Takeya[3], Yoshihiko Takano[3], Masashi Arita[4], Kenya Shimada[4], Hirofumi Namatame[4], Masaki Taniguchi[4], Katsuhiro Suzuki[5], Hidetomo Usui[6], Kazuhiko Kuroki[6], Takanori Wakita[1,2], Yuji Muraoka[1,2], and Takayoshi Yokoya[1,2**]

[1]*The Graduate School of Natural Science and Technology and Research Laboratory for Surface Science, Okayama University, Okayama 700-8530, Japan*

[2]*Research Center of New Functional Materials for Energy Production, Storage, and Transport, Okayama University, Okayama 700-8530, Japan*

[3]*MANA, National Institute for Materials Science, 1-2-1 Sengen, Tsukuba, Ibaraki 305-0047, Japan*

[4]*Hiroshima Synchrotron Radiation Center, Hiroshima University, Higashi-Hiroshima, Hiroshima 739-0046, Japan*

[5]*Research Organization of Science and Technology, Ritsumeikan University, 1-1-1 Noji-higashi, Kusatsu, Shiga 525-8577, Japan*

[6]*Department of Physics, Osaka University, 1-1 Machikaneyama, Toyonaka, Osaka 560-0043, Japan.*

(Dated: June 8, 2016)

Abstract

One of the ultimate goals of the study of iron-based superconductors is to identify the common feature that produces the high critical temperature ($T_c$). In the early days, based on a weak-coupling viewpoint, the nesting between hole- and electron-like Fermi surfaces (FSs) leading to the so-called $s\pm$ state was considered to be one such key feature. However, this theory has faced a serious challenge ever since the discovery of alkali-metal-doped FeSe (AFS) superconductors, in which only electron-like FSs with a nodeless superconducting gap are observed. Several theories have been proposed, but a consistent understanding is yet to be achieved. Here we show experimentally that a hole-like band exists in $K_xFe_{2-y}Se_2$, which presumably forms a hole-like Fermi surface. The present study suggests that AFS can be categorized in the same group as iron arsenides with both hole- and electron-like FSs present. This result provides a foundation for a comprehensive understanding of the superconductivity in iron-based superconductors.

Band-structure calculations predict that iron arsenide superconductors have hole- and electron-like Fermi surfaces (FSs) at the Brillouin zone (BZ) center and corner, respectively[1] [Fig. 1(a)]. These predictions have been confirmed by angle-resolved photoemission spectroscopy (ARPES),[2] which can directly elucidate FSs and determine the electron or hole character of FS sheets. Theoretical studies based on a weak-coupling approach have proposed that spin fluctuation arising from FS nesting induces the $s\pm$ superconducting state, where the gap function has $s$-wave symmetry with its sign reversed between the hole- and electron-like FSs.[3,4] This is consistent with the nearly isotropic $s$-wave superconducting gap in ARPES[2] and the spin resonance mode at the nesting vector observed in inelastic neutron scattering (INS) measurements.[5]

Unlike the iron arsenides, ARPES studies of alkali-metal-doped FeSe (AFS) superconductors have clarified that only electron-like FSs exist[6-16] [Fig. 1(b)] with a nodeless superconducting gap.[6-10] Theoretically, a $d$-wave superconducting state has been predicted for systems with only electron-like FSs.[4,17-19] This can explain INS results showing a spin resonance mode.[20] However, this was pointed out to be inconsistent with the absence of nodes on the electron FSs around the M point of the BZ.[21] Moreover, the $d$-wave is not consistent with the nodeless superconducting gap on the electron-like FS around the zone center.[6-8] To consistently explain the ARPES and INS observations, novel pairing states have been proposed,[21,22] such as a bonding-antibonding $s\pm$ superconducting state, where the gap function has $s$-wave symmetry with sign reversal between the two hybridized electron-like FSs at the M point. However, a recent theoretical study has shown that the $d$-wave dominates the novel $s$-wave for the FS topology observed by ARPES.[17] Hence, weak-coupling theories continue to face difficulties. On the other hand, the strong-coupling approaches based on a localized spin picture, which can explain part of the observations, also face difficulties. The strong-coupling theories predict the $s$-wave superconducting state for AFS,[23-25] which is consistent with a fully gapped state but inconsistent with the observation of the spin resonance mode.[20] The situation is the same for orbital-fluctuation-mediated pairing theories, which predict sign-conserving $s$-wave states.[26]

Thus, the proposed models based on both the weak- and strong-coupling approaches appear to fail to explain all the experimental results for AFS. In order to resolve this issue, revisiting the electronic structure of superconducting AFS is necessary. In $K_xFe_{2-y}Se_2$, it is known that a minority superconducting phase (~10%) and a majority insulating phase (~90%) coexist due to intrinsic phase separation.[27,28] Very recently, an increased area of the superconducting phase (~30%) has been found in a $K_{0.62}Fe_{1.7}Se_2$ single crystal obtained by a one-step method with the quenching technique.[29] Thus, there is a possibility that the ARPES measurement of $K_{0.62}Fe_{1.7}Se_2$ will detect an intrinsic electronic structure responsible for the superconductivity, which has eluded previous ARPES experiments because of the smaller superconducting region in the sample surface. In this article, we report the surprising observation of a hole-like band that probably forms the hole-like FS around the Γ point by performing ARPES on a high-quality $K_{0.62}Fe_{1.7}Se_2$ single crystal with $T_c$ = 32 K.[29,30]

High-quality $K_{0.62}Fe_{1.7}Se_2$ single crystals ($T_c$ = 32 K) were grown by the one-step method with the quenching technique,[29,30] where the quenching temperature was 700 °C. The chemical composition ratio was determined using an energy-dispersive X-ray spectrometer (EDS). ARPES measurements were carried out at BL-9A of Hiroshima Synchrotron Radiation Center (HSRC), where the energy of the light was set to 23 eV. The total energy resolution was set to 15 meV. Samples were cleaved and measured in situ under a vacuum better than 2 × 10$^{-9}$ Pa. Calibration of $E_F$ for the sample was achieved using a gold reference.

In order to compare the experimental band structure with the results of theoretical studies, a first-principles band calculation was also performed using the VASP package.[31,32] The lattice parameter values were those determined experimentally in Ref. 33. Here, we adopted the GGA-PBEsol exchange correlation functional.[34] The wave functions were expanded by plane waves up to a cutoff energy of 550 eV and 1000 $k$-point meshes were used. A ten-orbital tight-binding model was derived from the first-principles band calculation exploiting the maximally localized Wannier orbitals.[35,36] The Wannier90 code was used for generating the Wannier orbitals.[37] Some modifications were made to the original band structure for a better correspondence with the experiment; the interlayer hoppings within the $d_{xy}$ orbital were all multiplied by a factor of 0.5 assuming the reduction of the three-dimensionality, most likely due to correlation effects, and also the on-site energy of the $d_{xz/yz}$ orbitals was shifted by -0.1 eV, again a tendency that is due to correlation effects.[38)] The nearest-neighbor hopping of the $d_{xy}$ orbitals was also modified by -0.02 eV.

First, we demonstrate the low-lying electronic structure of $K_{0.62}Fe_{1.7}Se_2$ along Γ-M. Figures 2(a)-2(f) show the ARPES data taken along #1 in Fig. 2(q) with s-polarized (s-pol) and p-polarized (p-pol) light [for the geometrical measurement configuration, see Fig. 2(p)]. Around the Γ point, we identified six bands near $E_F$. In the s-pol data [Figs. 2(a)-2(c)], a large electron pocket (β) and two hole-like bands (δ, ε) are seen. In the p-pol data [Figs. 2(d)-2(f)], we observed a small electron pocket (α) and a hole-like band (γ). In addition, we also observed a faint intensity showing a finite dispersion (ζ) around $k = -0.5$ Å$^{-1}$ in Fig. 2(d). From the second-derivative method in energy and momentum distribution curve (MDC) analysis,[39)] we find that the ζ band has a faster hole-like dispersion than the ε band. This suggests that the ε and ζ bands are different. Figures 2(g)-2(l) show the polarization-dependent ARPES data along #2 in Fig. 2(q), corresponding to Γ-X. In these data, we observed the α-δ bands. In the p-pol data along Γ-X [Figs. 2(j)-2(l)], we found another $E_F$-approaching hole-like band that is dispersed from $E-E_F = -120$ meV and $k = -0.52$ Å$^{-1}$ to $E-E_F = -60$ meV and $k = -0.28$ Å$^{-1}$. The top of this hole-like band does not correspond to that of the ε band, which is located at $E-E_F = -60$ meV. In addition, the slope of this band in a certain k-region (-0.52 Å$^{-1}$ < k < -0.38 Å$^{-1}$) is similar to that of the ζ band in Fig. 2(d), and is three times larger than that of the ε band in Fig. 2(a). These results indicate that the $E_F$-approaching hole-like band in Figs. 2(j)-2(l) is the ζ band. Note that the ε band with a narrow dispersion has almost no intensity in the polarization-dependent ARPES data along Γ-X. Around the M point, four bands can be derived from the ARPES data taken along #3 in Fig. 2(q) with circular polarized (c-pol) light [Figs. 2(m)-2(o)]. Near $E_F$, we observed shallower and deeper electron-like pockets that are degenerate around -10 meV. In the higher-binding-energy region (-200 meV < $E - E_F$ < -120 meV), two different dispersions are seen in the second-derivative plot and EDCs of Fig. 2(m) [Figs. 2(n) and 2(o)]: a hole-like dispersion (-0.56 Å$^{-1}$ < k < 0.4 Å$^{-1}$) and an electron-like dispersion (k > 0.4 Å$^{-1}$). Figure 2(r) shows a summary of the band structure along the Γ-M direction derived from Figs. 2(a)-2(f) and 2(m)-2(o). We now find that the δ and ζ bands are connected to the hole-like and electron-like dispersions around M, respectively. Except for the ζ band, the observed bands are found to be consistent with previous ARPES studies[6-16)]. The newly identified ζ band with a steep hole-like dispersion approaches $E_F$ and possibly crosses $E_F$ around Γ, although the near-$E_F$ dispersion of the ζ band along Γ-M and Γ-X is unclear owing to its weak intensity as can be seen in Figs. 2(d) and 2(j).

In order to elucidate the near-$E_F$ dispersion of the ζ band, we measured the ARPES spectra along several cuts [#3-#5 in Figs. 3(a) and 3(c)] with c-pol light. Along #3 and #4 [Figs. 3(d) and 3(e), respectively], we observed an electron-like dispersion near $E_F$, corresponding to the β band. The FS and the constant-energy surface of the β band can be seen in the energy contour plots at 0 and -20 meV, respectively [Figs. 3(a) and 3(b), respectively]. While the intensity of the β band is almost completely suppressed along #5, possibly due to the matrix element effects, a fast hole-like dispersion, corresponding to the ζ band, is seen around $k = +0.3$ Å$^{-1}$ in Figs. 3(f) and 3(g). A large constant-energy surface of the ζ band can be seen in the energy contour plot at -40 meV, as indicated with a red broken circle in Fig. 3(c). Comparing the dispersion of the ζ band at #1 and #5 [Fig. 3(h)], we find that the ζ band gradually shifts to a lower-momentum region as the wave vectors move away from the Γ-M line, demonstrating the presence of a hole-like ζ band around Γ. Although the intensity of the ζ band almost vanishes in the energy contour plot at -20 meV, the MDCs of the ARPES plot along #5 [Fig. 3(g)] show a finite energy dispersion of the shoulder structure that reaches -20 meV at $k = +0.26$ Å$^{-1}$, suggesting that $E_F$ crosses the ζ band. As mentioned above, the ζ band loses its intensity near $E_F$, and this is discussed later in connection with theoretical studies.

Now let us compare our experimental observations with the theoretical band structure. We have obtained a ten-orbital model of $KFe_2Se_2$ from first-principles calculation.[31,32,37] In Fig. 4(a), we present its band structure along the Γ-M line with some modifications (as described before), together with the orbital-resolved band structures in Figs. 4(b)-4(d). For the p-pol light along Γ-M, bands with $d_{xz}$, $d_{z2}$, and $d_{x2-y2}$ orbital characters can be detected (regarding the selection rule for d orbitals, see Ref. 40), but since $d_{z2}$ and $d_{x2-y2}$ orbital weights are not expected to be present around $E_F$ in $KFe_2Se_2$, only the $d_{xz}$ orbital should be observed. In Fig. 4(c), there is indeed a band whose $d_{xz}$ orbital character is relatively strong in the high-binding-energy regime that becomes weaker near $E_F$. This perfectly matches with the observed intensity variation of the newly observed ζ band in the p-pol data given in Figs. 2(d)-2(f). Theoretically, this hole-like band is often referred to as the "$d_{xy}$ band" because it has dominant $d_{xy}$ character, especially near $E_F$ [see Fig. 4(d)], but it is highly likely that we have experimentally detected this "hidden band" for the first time by capturing its $d_{xz}$ component. Some possible reasons why the $d_{xy}$ component of this hole-like band was not observed for the *s*-pol light [Figs. 2(a)-2(c)] will be discussed below.

Regarding the bands other than the ζ band, we found that the experimentally observed α, γ, and δ bands can be assigned to the theoretically predicted $d_{xz}$ electron-like, $d_{xz}$ inner hole-like, and $d_{yz}$ middle hole-like bands, respectively. Considering the selection rule for *d* orbitals,[40] the assignment of these bands is found to be consistent with their polarization dependence along #1 [Figs. 2(a) and 2(d)]. The γ ($d_{xz}$) and δ ($d_{yz}$) bands are renormalized by a factor of ~2 compared with the calculation result, indicating the presence of the electron correlation effect for $d_{xz/yz}$ bands. Regarding the correlation effect of the ζ (hybridized $d_{xy}$-$d_{xz}$) band, the renormalization strength is found to be similar to that of the γ ($d_{xz}$) and δ ($d_{yz}$) bands (a factor of ~2), as can be seen from the fact that both in the experiment and the theory, they have about the same gradient. Around the M point, the correspondence between the experiment [Fig. 2(r)] and the calculation suggests that the shallower and deeper electron-like bands correspond to the $d_{xz/yz}$ and $d_{xy}$ bands, respectively. Thus, the $d_{xy}$ orbital components around both Γ and M show almost no spectral intensity. Several possibilities can be considered as the origin for this suppressed intensity: strong correlation,[12] impurity scattering,[41] or matrix element effects peculiar to the $d_{xy}$ orbital, especially around Γ.[42]

Theoretically, the position of the hole-like ζ and electron-like α bands is strongly affected by the relation between the nearest-neighbor ($t_1$) and next-nearest-neighbor ($t_2$) hoppings within the $d_{xy}$ orbital.[43] In Ref. 44, some of the present authors have shown that a peculiar relation $t_1 < t_2$ enlarges the overlap between the two bands and also enhances spin-fluctuation-mediated s± superconductivity. For the model of KFe$_2$Se$_2$, $t_1$ = -0.008 eV and $t_2$=0.059 eV are obtained, namely $t_1 \ll t_2$. Hence, the observation of both the hole-like (ζ) and electron-like (α) FSs is consistent with the theoretical expectation.

The experimentally observed β and ε bands are not predicted in our band calculations, although the ε band was assigned as the $d_{xy}$ band and its strong renormalization was discussed in terms of the orbital-selective Mott phase (OSMP).[12] This means that the number of the observed bands [Fig. 2(r)] is larger than that in the band calculations. The presence of surface-related bands is one possible explanation for the difference in the number of bands. Another possibility is that the intrinsic phase separation may induce different metallic phases in $K_xFe_{2-y}Se_2$. A recent scanning micro-X-ray diffraction study has revealed the existence of an interface phase that surrounds and protects the filamentary network of the metallic phase embedded in the insulating phase.[45] Reference 45 suggests that the interface phase is likely to be the OSMP, where the $d_{xy}$ bands are specifically localized while the other bands are itinerant.[12] The interface phase may induce the bands that are not predicted by the band calculation.

The observation of the hidden hole-like band approaching $E_F$ suggests the presence of a hole-like FS in $K_xFe_{2-y}Se_2$. This result indicates that AFS can be categorized in the same group as iron arsenides with both hole- and electron-like FSs present. Thus, the "common identity" of the iron-based superconductors may be the presence of hole- and electron-like FSs. In order to confirm this indication, we suggest an experimental investigation of whether the hole-like FS exists in single-layer FeSe films, which are believed to be superconducting below $T_c \sim 60$ K.[46-48]

**Acknowledgements**

The polarization dependent ARPES experiments at HSRC were performed with the approval of HSRC (Proposal Nos. 14-A-21 and 15-A-15). The preliminary ARPES measurements were performed at the Photon Factory and SPring-8 under proposal numbers 2011G086 and 2011B1892, respectively. M.S. was supported by a Grant-in-Aid for JSPS Fellows. This work was also partially supported by Grants-in-Aid for Scientific Research on Innovative Areas "Heavy Electron" (No.20102003), for Scientific Research (B) (No.24340079), for Young Scientists (B) (No.25800205), and also for JSPS Fellows (No.25009605) from the Ministry of Education, Culture, Sports, Science and Technology of Japan (MEXT). This work was also partially supported by the Program for Promoting the Enhancement of Research Universities from MEXT.

*sc20217@s.okayama-u.ac.jp, **yokoya@cc.okayama-u.ac.jp

**Figure Captions**

Fig. 1. (Color online) (a,b) Schematic Fermi surface topology of iron arsenides and alkali-doped iron selenides, respectively. The red and blue circles represent hole- and electron-like FSs, respectively.

Fig. 2. (Color online) (a) ARPES intensity plot taken along #1 with *s*-pol light. (b,c) Second derivatives with respect to energy and the EDCs of (a), respectively. (d-f) Same as (a-c) but taken along #1 with *p*-pol light. (g-l) Same as (a-f), but taken along #2. (m-o) Same as (a-c) but taken along #3 with c-pol light. In these data, open and filled red circles represent the peak positions determined from analyses of the EDCs and MDCs, respectively, and white lines are guides for the eye. All data are taken at 7 K (below $T_c$) except (g-l), which are taken at 40 K (above $T_c$) (p) Experimental geometry for polarization-dependent ARPES measurements. The *x* and *y* axes are along the Γ-M

line in (q). (q) Two-dimensional BZ (black line) and the measurement directions (orange arrows). (r) Summary of the experimental band structure for $K_{0.62}Fe_{1.7}Se_2$. The red, blue, and green open (filled) circles correspond to the EDC (MDC) peaks in (a), (d), and (m), respectively. Open and filled squares denote peak positions obtained by applying the mirror symmetry operation to the original peak positions [circles in (r)] with respect to the Γ and M points. The black and red lines represent the band dispersions deduced from the present ARPES study.

Fig. 3. (Color online) (a-c) Energy contour intensity plots at $E-E_F$ = 0, -20, and -40 meV, respectively. In (a-c), solid and broken circles indicate the FSs and the constant-energy contours deduced from the MDC peak positions (red and blue dots). (d-f) ARPES intensity plots taken along #3-#5, respectively, as indicated by the white lines in (a) and (c). (g) MDCs of (f) together with the peak position (red circles). (h) Dispersion of the ζ band at #1 (red) and #5 (green). In Fig. 3, all data were taken at $hv$ = 23 eV with c-pol light.

Fig. 4. (a) Band structure of the ten-orbital model of $KFe_2Se_2$ along Γ-M. (b) Contribution of the $d_{yz}$ orbital to the band dispersion. (c,d) Same as (b) but for the $d_{xz}$ orbital and $d_{xy}$ orbital, respectively.

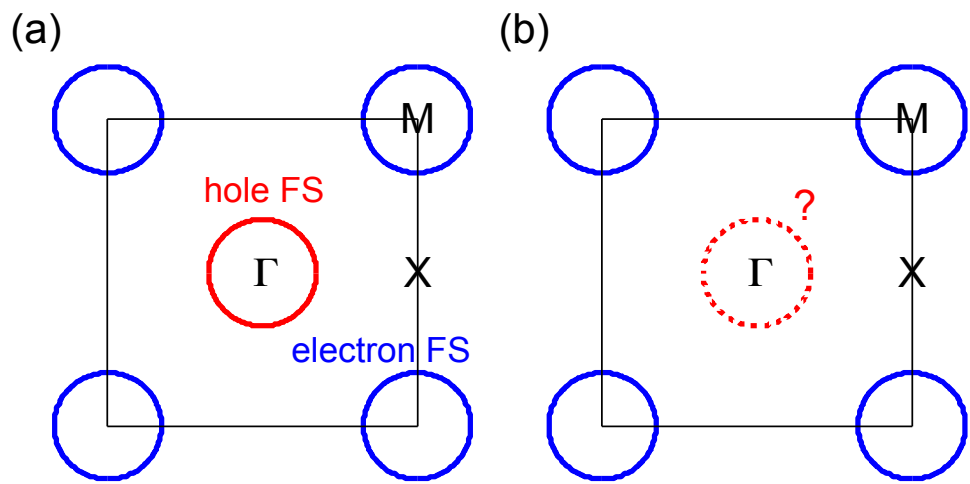

Fig. 1 M. Sunagawa et al.

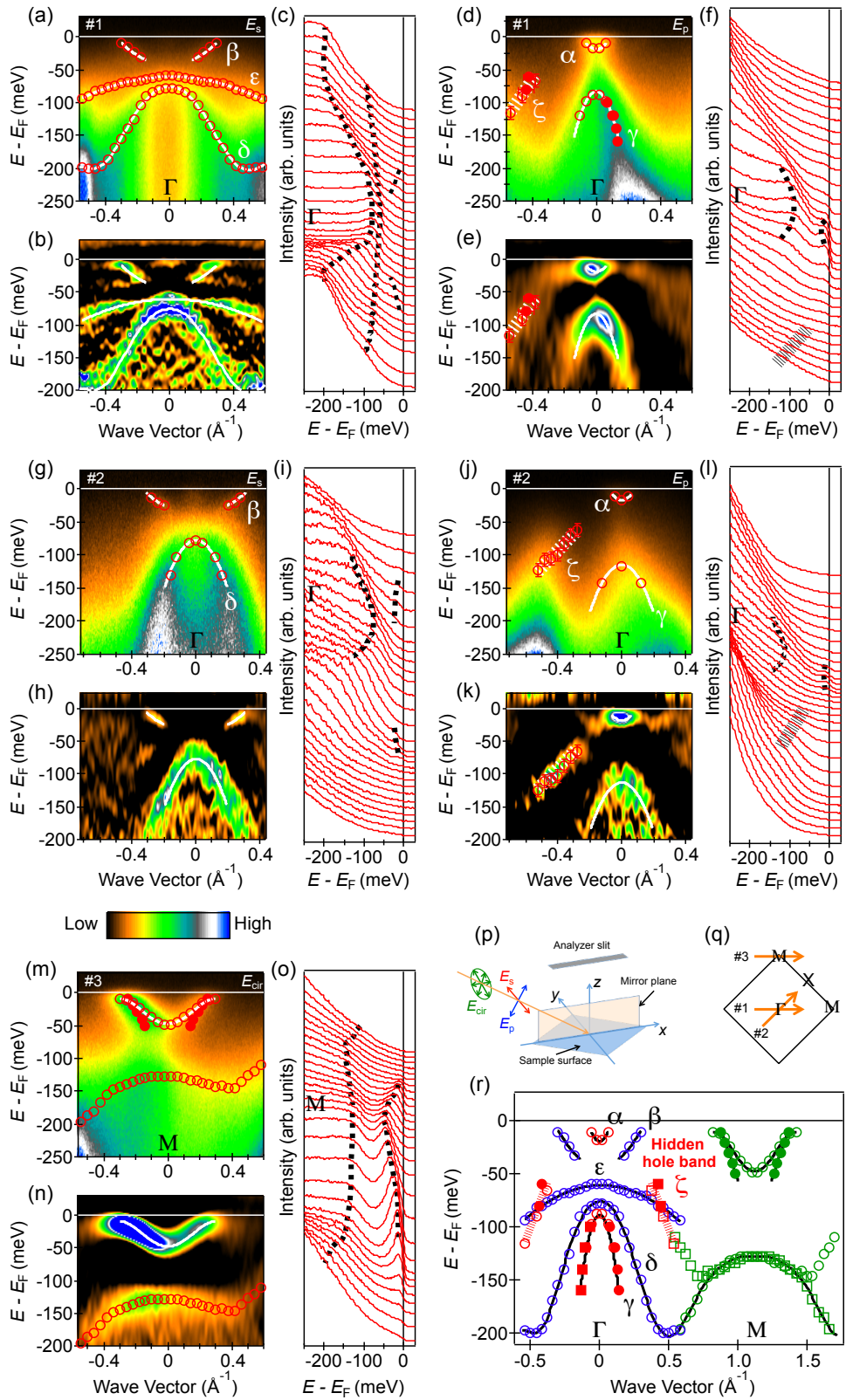

Fig. 2 M. Sunagawa et al.

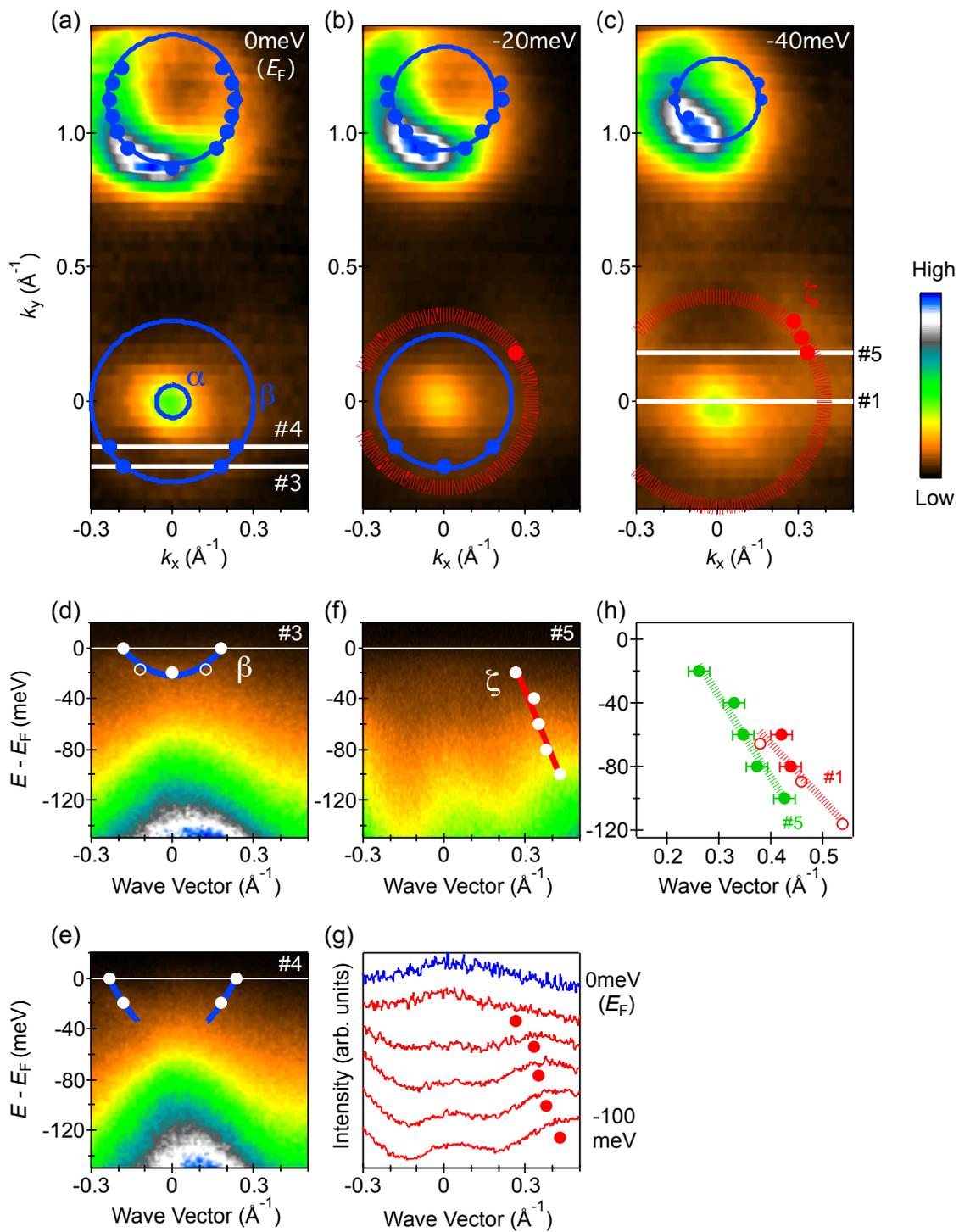

Fig. 3 M. Sunagawa et al.

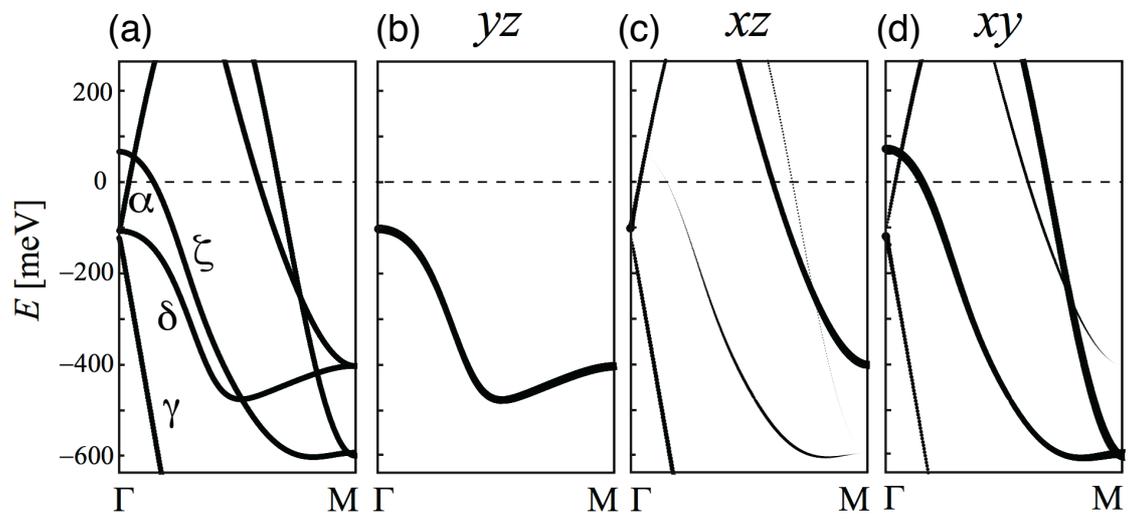

Fig. 4 M. Sunagawa et al.

# Supplementary Information

## Supplementary Figures

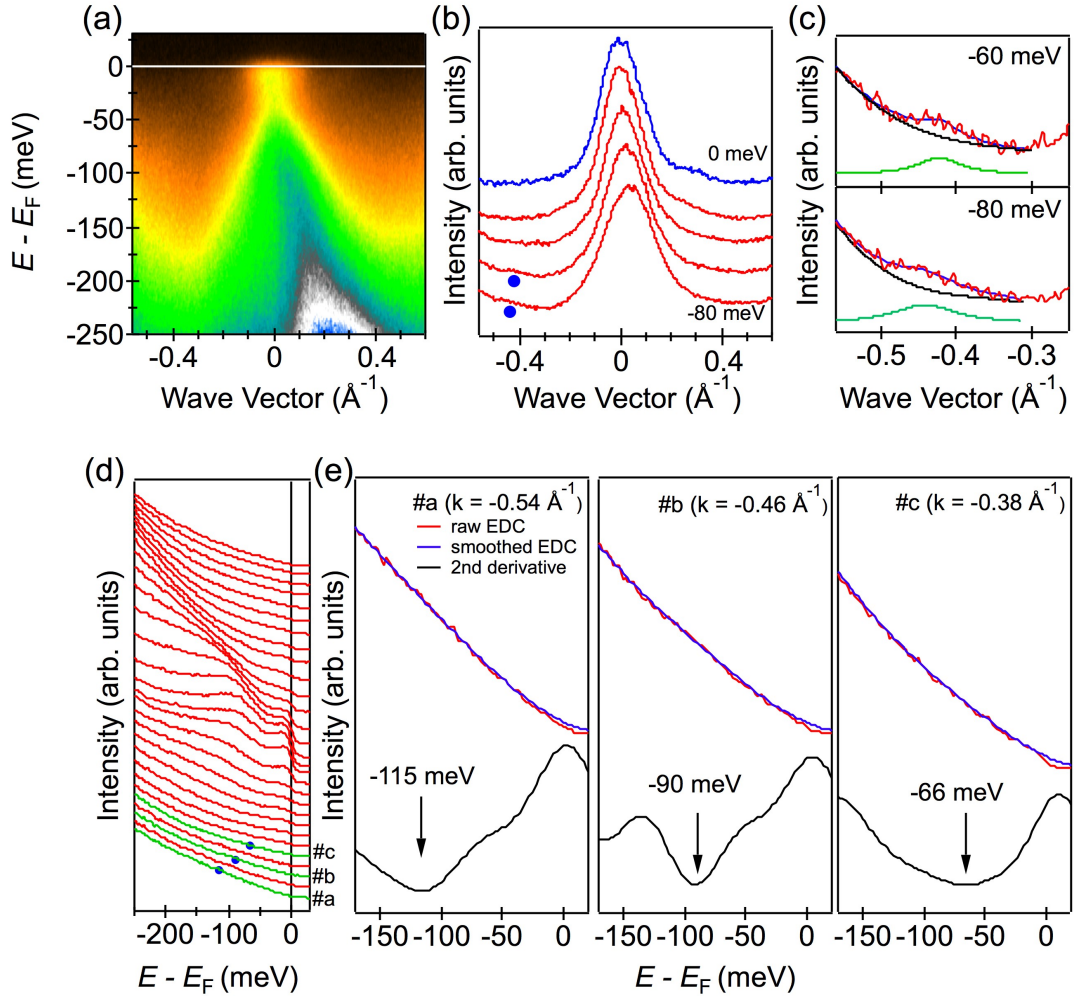

Supplemental Figure S1  MDC and EDC analyses of *p*-pol ARPES data. (a,b) The *p*-pol ARPES plot taken along #1 in Fig. 2(q) of the main text and its MDCs, respectively. In (b), the blue dots represent the peak position, corresponding to the ζ band. (c) The MDCs at -60 meV and -80 meV (red lines). In c, the blue line is a sum of the Lorentzian of the ζ band (green line) and the background (black line). (d) EDC of (a). (e) The selected EDCs (red line), together with the smoothed EDC (blue line) and its second derivative (black line). In (e), The black arrows mark the peak position. Note that the EDCs corresponding to #a, #b, #c are drawn as green lines in (d).

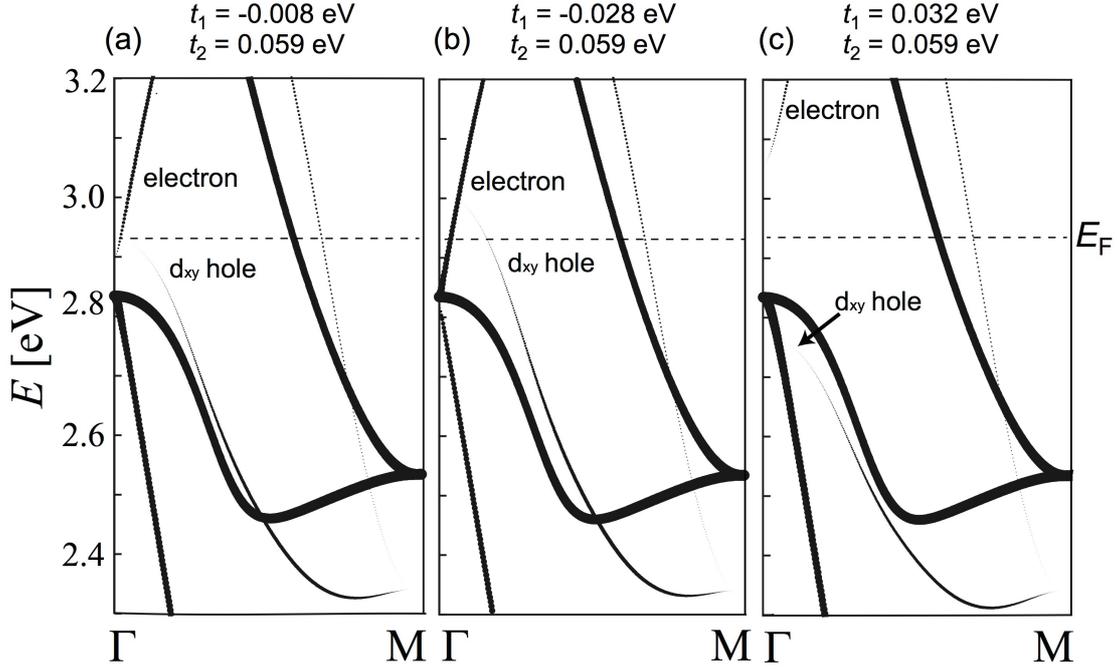

Supplemental Figure S2    Calculated band structure for various $t_1$. (a) Band structure of the ten orbital model of $KFe_2Se_2$ along Γ-M. The thickness of the lines represents the strength of the $d_{xz/yz}$ orbital character, so that the $d_{xy}$ bands are thin. The interplane hoppings within the $d_{xy}$ orbital is multiplied by a factor of 0.5, and the on-site energy of the $d_{xz/yz}$ orbitals is shifted by -0.1 eV. Here, the nearest neighbor ($t_1$) and the next nearest neighbor ($t_2$) hopping within the $d_{xy}$ orbital $t_1$ = -0.008 eV and $t_2$ = 0.059 eV, respectively (the original value obtained from first principles calculation[1-3]). (b) same as (a) except $t_1$ is decreased by 0.02 eV ($t_1$ = -0.028). (c) same as (a) except $t_1$ is increased by 0.04 eV ($t_1$ = 0.032). The correspondence with the experimental result becomes better in (b), suggesting that $t_1$ may be even further decreased by effects beyond the band calculation. In any case, the electron-like bands crossing $E_F$ around Γ also observed in previous studies,[4-11] is a strong indication of small (or negative) $t_1$, which in turn is consistent with the observation of the hidden hole-like ζ band with approaching $E_F$.